# Edge-based Tensor prediction via graph neural networks

Yang Zhong[1,2#], Hongyu Yu[1,2#], Xingao Gong[1,2], Hongjun Xiang[1-3*]

[1]Key Laboratory of Computational Physical Sciences (Ministry of Education), State Key Laboratory of Surface Physics, and Department of Physics, Fudan University, Shanghai, 200433, China
[2]Shanghai Qizhi Institution, Shanghai, 200030, China
[3]Collaborative Innovation Center of Advanced Microstructures, Nanjing, 210093, China
*E-mail: hxiang@fudan.edu.cn
#These authors contribute equally to this work.

## Abstract

Message-passing neural networks (MPNN) have shown extremely high efficiency and accuracy in predicting the physical properties of molecules and crystals, and are expected to become the next-generation material simulation tool after the density functional theory (DFT). However, there is currently a lack of a general MPNN framework for directly predicting the tensor properties of the crystals. In this work, a general framework for the prediction of tensor properties was proposed: the tensor property of a crystal can be decomposed into the average of the tensor contributions of all the atoms in the crystal, and the tensor contribution of each atom can be expanded as the sum of the tensor projections in the directions of the edges connecting the atoms. On this basis, the edge-based expansions of force vectors, Born effective charges (BECs), dielectric (DL) and piezoelectric (PZ) tensors were proposed. These expansions are rotationally equivariant, while the coefficients in these tensor expansions are rotationally invariant scalars which are similar to physical quantities such as formation energy and band gap. The advantage of this tensor prediction framework is that it does not require the network itself to be equivariant. Therefore, in this work, we directly designed the *edge-based tensor prediction graph neural network* (ETGNN) model on the basis of the invariant graph neural network to predict tensors. The validity and high precision of this tensor prediction framework were shown by the tests of ETGNN on the extended systems, random perturbed structures and JARVIS-DFT datasets. This tensor prediction framework is general for nearly all the GNNs and can achieve higher accuracy with more advanced GNNs in the future.

## I. INTRODUCTION

So far, density functional theory (DFT) is still a mainstream theoretical method for calculating material properties. In principle, most of the physical properties of a material can be known by solving the Schrödinger equation of the system. However, the huge computational cost of solving the Schrödinger equation makes it difficult to screen a large number of materials quickly. Similar to DFT, machine learning (ML) can also build direct mappings from the structures to the properties with much faster speed. In recent years, ML has shown powerful ability in high-throughput screening of high-performance alloys[1], lead-free perovskite photovoltaic materials[2, 3], high-efficiency catalytic materials[4-6], advanced lithium batteries[7, 8] and high-$ZT$ thermoelectric materials[9]. Unfortunately, at present most ML methods can only build mappings from the structures to the scalar properties, which hinders the application of ML in predicting the tensor properties of the materials.

The deep neural networks for predicting material properties can mainly be divided into two types: one is the neural networks with manual features or so-called "descriptors" as input[10-13]. "Descriptors" are usually various physical and chemical properties of molecules or numerical expressions derived from various symmetry functions based on the molecular structures. Some typical neural networks based on "descriptors" are DPMD[14], CFID[15], FCHL[16], etc. The other is based on the framework of message passing neural networks (MPNNs)[17], which mainly includes various types of graph neural networks, such as DTNN[18], SchNet[19], DimeNet++[20], CGCNN[21], etc. MPNNs only need the input structure information of molecules and crystals, and learns the representation of each atom in its local environment through message passing between atoms. Since MPNNs can learn different representations according to the target properties, they can often achieve higher accuracy on the datasets containing different kinds of structures. In the present work, the tensor prediction framework is designed based on MPNN.

Standard MPNNs use the transformationally invariant messages to update the transformationally invariant node features. Standard invariant MPNNs can output some invariant physical quantities such as formation energy, band gap, heat capacity, etc. However, the standard MPNN cannot directly predict the tensor properties that rotate equivariantly with the spatial coordinate system. Schütt *et al.* recently has proposed an equivariant graph neural network called PAiNN[22], which is based on MPNN and has introduced additional vectorial node features. These vectorial features are updated by equivariant message passing functions constructed by linear operations. Theoretically, PAiNN can construct any order equivariant tensor outputs from the 1st order vectorial node features. However, PAiNN showed poor performance in fitting a tensor regression model for the BECs in our own tests. As far as we know there is still a lack of a reliable tensor prediction framework that can be implemented on graph neural networks.

In this work, we proposed a tensor prediction framework on graph neural networks. The main idea of this framework is that the tensor property of a crystal can be represented as the average of the tensor contributions of all the atoms in the crystal, and that the tensor contribution of each atom can be expanded as the sum of the tensor projections in the directions of the edges connecting the atoms. In order to implement this framework, we designed *edge-based tensor prediction graph neural network* (ETGNN) and tested its tensor prediction ability for the forces, BECs and DL tensors. This approach is more intuitive and enables many invariant graph neural networks to be able to predict tensors without introducing extra vectorial features.

## II. THEORY

The standard MPNN decomposes the total energy of a molecule or crystal into the sum of the energy of each atom[23], and the energy of each atom is extracted from the final updated atom features. The atom features are updated through multiple message passing layers which simulate the interaction between the atoms. In this work, we have proposed a general tensor prediction framework that represents the crystal rank-$M$ tensor property ($T^M$) as the average of the atomic rank-$M$ tensor contributions ($T_j^M$) which can be decomposed into the sum of the tensor projections in the directions of the edges connecting the atom. The key to the problem is how to predict the atomic tensors. The illustration of predicting the atomic rank-*M* tensors in this framework is shown in Fig. 1. This tensor prediction framework can be expressed by the following formulas:

$$T^M = \frac{1}{N} \sum_{j=1}^{N} T_j^M, \qquad (1)$$

$$T_j^M = \sum_{k_i \in N(j)} t_{k_1 k_2 \cdots k_M j} \vec{e}_{k_1 j} \otimes \vec{e}_{k_2 j} \otimes \cdots \otimes \vec{e}_{k_M j}, \quad (2)$$

where $\vec{e}_{k_i j}$ is the unit edge vector from atom $k_i$ to atom $j$. $N(j)$ represents the set of all neighbor atoms of atom $j$ within a cutoff radius $r_c$. Eq. (2) is a general expression of the atomic rank-$M$ tensor expansion which requires at most $M$ individual edge vectors to construct the tensor bases. Since most of the crystal tensors $T^M$ are symmetric, then we can decompose $T^M$ into the average of the symmetric atomic tensors which can be expanded with the symmetric tensor bases. Let $n$ be the number of the necessary individual edge vectors required by the rank-$M$ tensor bases, i.e., $n$ is the cardinality of the set $\{k_i | 1 \leq i \leq M\}$. $n$ can be less than $M$ since the symmetric tensor bases can be constructed by the direct product of repeated edge vectors. This can greatly reduce the complexity of the formulas and the amount of unnecessary calculations and is conducive to the training of the networks. For the coefficient $t_{k_1 k_2 \cdots k_M j}$ containing the indexes of $n$ individual neighbors, it can be derived from the features $m_{k_1 k_2 \cdots k_M j}$ of the cluster with $n+1$ atoms around atom $j$. The equivariant tensors constructed by Eq. (1) and (2) are applicable to all periodic and nonperiodic systems. Below we introduce the expansion of the tensors with respect to the edge directions using force vector, BEC, DL and PZ tensor as examples.

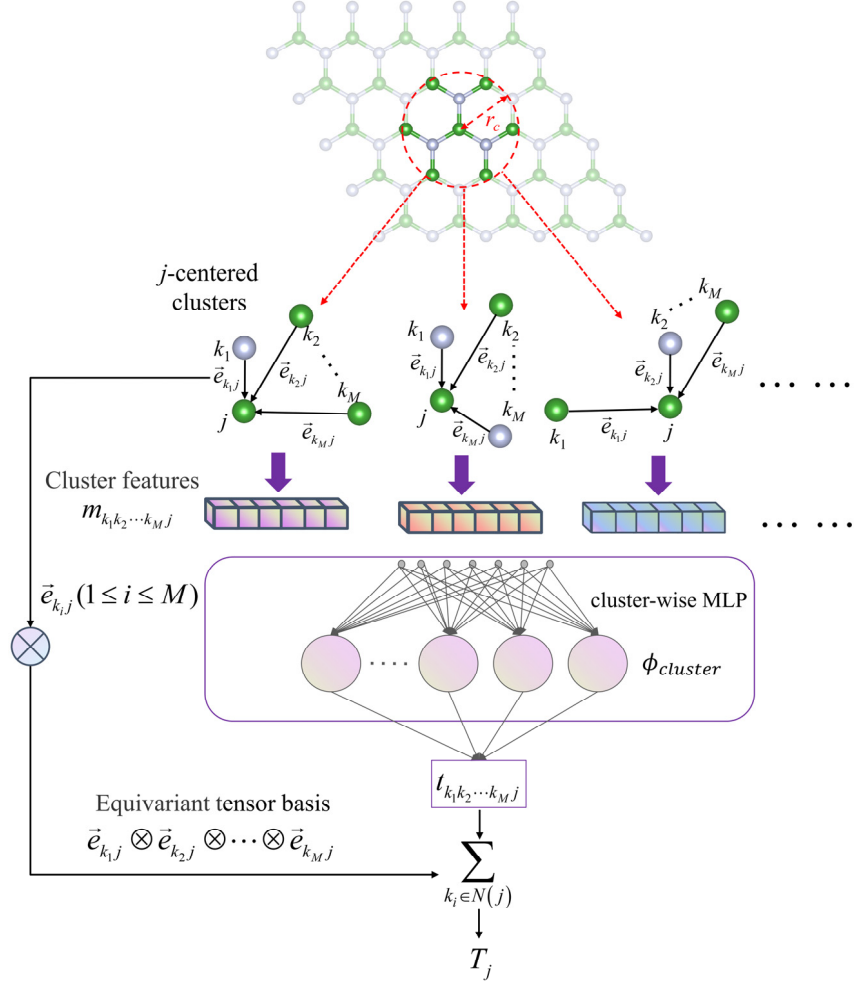

**Figure 1:** The schematic diagram of predicting the atomic rank-*M* tensors. The atomic rank-*M* tensors are expanded with the rank-*M* tensor bases constructed by *M* edge vectors which contain at most *M* individual neighbor atoms within a cutoff radius $r_c$. For symmetric tensor bases, the required number of individual neighbor atoms *n* can be less than *M* by using repeated edge vectors. The *n* individual neighbor atoms together with the central atom *j* form a cluster containing (*n*+1) atoms. The lengths of the edges and the angles between the edges in the cluster can be aggregated into a feature vector $m_{k_1 k_2 \cdots k_M j}$ that can be updated in MPNN. The corresponding tensor expansion coefficients $t_{k_1 k_2 \cdots k_M j}$ are mapped from the features of each cluster through a cluster-wise regression layer $\phi_{cluster}$. The tensor of atom *j* is obtained by summing the tensor contributions of each *j*-centered clusters.

## A. Equivariant expression of forces

The force on the $j$-th atom is the sum of the forces from the surrounding neighbor atoms:

$$\vec{F}_j = \sum_{k \in N(j)} t_{kj} \vec{e}_{kj}. \quad (3)$$

The coefficients $t_{kj}$ are the magnitude of the forces between the atoms. Due to the complex nonlinear many-body interactions between the atoms in the crystals, it is difficult to use an explicit formula to calculate $t_{kj}$. However, $t_{kj}$ can be learned directly through the powerful fitting ability of the deep neural networks. The edge $E_{kj}$ between atom pairs $kj$ can be used as a node in the line graph[24] to update its features with the surrounding environment. The coefficient $t_{kj}$ can be extracted from the final updated features $m_{kj}$ of the edge $E_{kj}$ through an edge-wise neural network: $t_{kj} = \phi_{edge}(m_{kj})$。Since the direction of the edge rotates with the crystal, it can be proved that Eq. (3) is rotationally equivariant, see appendix A.

**B. Equivariant expression of BECs**

Since BEC is an asymmetric second-order tensor, the terms in Eq. (2) are explicitly split into the symmetric part $T_j^{sym}$ and the non-symmetric part $T_j^{non-sym}$:

$$T_j = T_j^{sym} + T_j^{non-sym}, \tag{4}$$

$$T_j^{sym} = \sum_{k \in N(j)} t_{kj} \vec{e}_{kj} \otimes \vec{e}_{kj}, \tag{5}$$

$$T_j^{non-sym} = \sum_{\substack{k,i \in N(j) \\ k \neq i}} t_{kji} \vec{e}_{kj} \otimes \vec{e}_{ji}. \tag{6}$$

$T_j^{sym}$ is expanded with the second-order symmetric tensor bases $\vec{e}_{kj} \otimes \vec{e}_{kj}$ constructed by the unit direction vectors $\vec{e}_{kj}$ between the atom pairs $kj$. $T_j^{non-sym}$ is expanded with the asymmetric tensor bases $\vec{e}_{kj} \otimes \vec{e}_{ji}$ constructed by the unit direction vectors $\vec{e}_{kj}$ and $\vec{e}_{ji}$. Since $T_j^{sym}$ and $T_j^{non-sym}$ require one and two individual neighbor atoms respectively, the expansion coefficients $t_{kj}$ and $t_{kji}$ are mapped from the hidden features of the edge messages ($m_{kj}$) and the triplet messages ($m_{kji}$) respectively. The embedding and updating of the triplet messages ($m_{kji}$) will be illustrated in detail in the following network implementation section. It is important to note that the expansion formula of BEC applies to all second order atomic tensors. It can be proved by linear algebra that the expression of the BEC is rotationally equivariant, see appendix A.

**C. Equivariant expression of DL tensors**

The DL tensors are second-order symmetric tensors of the crystals and are not extended with the expansion of the primitive cells of the crystals. The DL tensors can be represented as the average of the atomic contributions of the second-order symmetric tensors $T_j^{sym}$. Therefore, the edge-based expansion of the DL tensor can be written as:

$$DL = \frac{1}{N}\sum_{j=1}^{N}\sum_{k \in N(j)} t_{kj}\vec{e}_{kj} \otimes \vec{e}_{kj} . \tag{7}$$

In addition, since the second-order tensor bases are equivariant, the DL tensor calculated in this way is also equivariant.

### D. Equivariant expression of piezoelectric tensors

The PZ tensor is a third-order symmetric tensor. According to the above discussion, the edge-based expansion formula of the PZ tensor is given as follows:

$$PZ = \frac{1}{N}\sum_{j=1}^{N}\left(\sum_{k \in N(j)} t_{kj}\vec{e}_{kj} \otimes \vec{e}_{kj} \otimes \vec{e}_{kj} + \sum_{\substack{k,i \in N(j) \\ k \neq i}} t_{kji}\vec{e}_{kj} \otimes \vec{e}_{ji} \otimes \vec{e}_{ji}\right) \tag{8}$$

This expression is symmetric and rotationally equivariant.

### III. NETWORK IMPLEMENTATION

In traditional GNNs, the features of each atom are embedded in a high-dimensional space and updated by passing messages between atoms. The contribution of each atom to the scalar property of a crystal is finally obtained through an atom-wise regression neural network. According to the tensor prediction framework discussed in the previous section, the expansion coefficients of tensors up to order three are mapped from the features of the edges or the triplets. In this work, we designed the edge-based tensor prediction graph neural network (ETGNN), which is a modified version of DimeNet++[20]. DimeNet++ is an invariant graph neural network focusing on scalar regression problems. ETGNN obtains the tensor expansion coefficients by updating the learned embeddings of the edges and the triplets in the crystal graphs.

The graph representation of the crystals in the traditional crystal graph neural network consists of two parts: (1) the node feature vectors $h_i$ that represent the attributes of the atoms and (2) the edge feature vectors $u_{ij}^{(k)}$ that represent the attributes of the interaction between the adjacent atoms. Since the periodicity of the crystals, the crystal graph is a multigraph with multiple edges between two adjacent atoms in the primitive cell. *k* is the index of the multiplicity of the edges. The following mapping is realized in the traditional crystal graph neural networks:

$$f_\theta : \{h_i, \cdots, h_j, \cdots u_{ij}^{(k)}, \cdots\} \to R, \tag{9}$$

where $\theta$ is the learnable parameters of the networks.

ETGNN inherits the directional message passing framework used by DimeNet++, i.e. the features of atom pairs *ij* and their edges $u_{ij}^{(k)}$ are integrated into the directional edge message $m_{ji}^{(k)}$ being sent from atom *j* to atom *i*. In addition, ETGNN also introduces an additional triplet component in the crystal graph, which is composed of node *j* and its two neighbor nodes *k* and *i*. The nodes, edges, and angles contained in the triplets are embedded into the triplet messages $m_{kji}$. Using equivariant expressions of tensors proposed in the present work, ETGNN can predict both scalars and various tensor properties, such as forces $\vec{F}$, BECs, DL and PZ tensors. ETGNN can realize the following mapping:

$$f_\theta : \{m_{kj}^{(k_1)}, \cdots, m_{ji}^{(k_2)}, \cdots, m_{kji}^{(k_1,k_2)}, \cdots\} \to R, \vec{F}, \text{BEC}, \text{DL}, \text{PZ}, \cdots \tag{10}$$

where $\theta$ is the learnable parameters of the network.

The architecture of ETGNN is shown in Fig. 2(a). Based on the DimeNet++ framework, ETGNN has introduced the triplet embedding block and the triplet update block. For tensor prediction, an edge-wise MLP $\phi_{edge}$ and a triplet-wise MLP $\phi_{triplet}$ are used to get the coefficients in the tensor expansion: $t_{kj} = \phi_{edge}(m_{kj})$, $t_{kji} = \phi_{triplet}(m_{kji})$.

**Embedding block.** The radial Bessel function $\phi_{RBF} : R^1 \to R^{nb}$ is used to expand the distance between atom *i* and its neighbor *j* within a cutoff radius $r_c$:

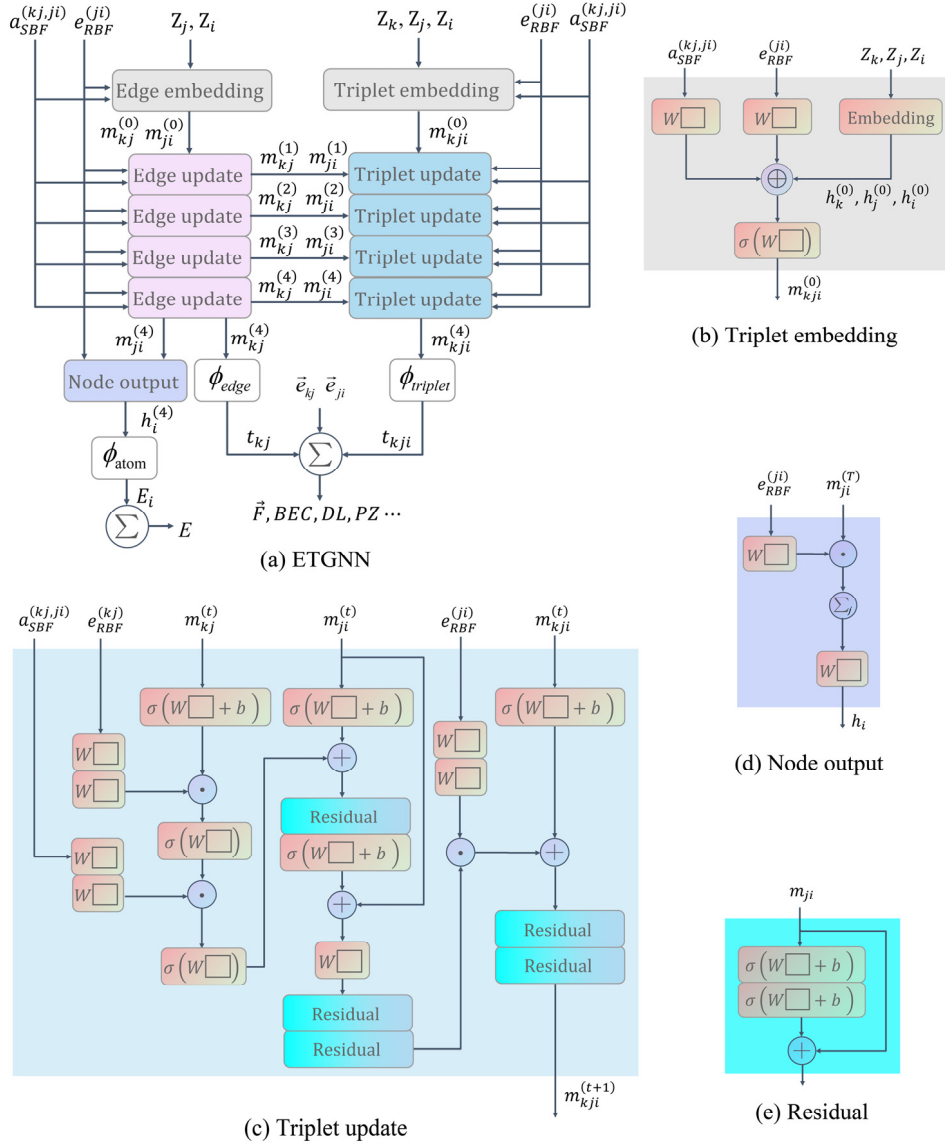

**Figure 2**: The architecture of ETGNN and its sub-networks. (a) is the architecture of ETGNN with four layers of edge/triplet update blocks. The edge embedding and edge update blocks are inherited from the original embedding and interaction blocks of DimeNet++ respectively, so that the high precision of DimeNet++ can be retained. The initial features of the triplets are generated in (b) the triplet embedding block and are updated through (c) the triplet update blocks. The output of the last edge update block can be aggregated into the node features for scalar outputs by the (d) node output block. (e) is the residual block[25], which is a multilayer feedforward neural network with a shortcut connection to avoid the vanishing gradients problem.

$$\phi_{RBF}\left(\left|\vec{r}_{ji}\right|\right) = \sqrt{\frac{2}{r_c}} \frac{\sin\left(n\pi\left|\vec{r}_{ji}\right|/r_c\right)}{\left|\vec{r}_{ji}\right|}, \tag{11}$$

where $\left|\vec{r}_{ji}\right|$ represents the distance between atom pairs $ij$, and $n \in \{1,\cdots,nb\}$. $nb$ is the number of radial basis functions, $nb = 8$ in this work. The following cosine cutoff function is used for the distance expansion coefficients:

$$f_c(|\vec{r}_{ji}|) = \begin{cases} 0.5 \cdot \left[\cos\left(\pi |\vec{r}_{ji}|/r_c\right) + 1\right] & \text{for } |\vec{r}_{ji}| \leq r_c \\ 0 & \text{for } |\vec{r}_{ji}| > r_c \end{cases}, \quad (12)$$

which can ensure continuous behavior when the atoms enter or leave the cut-off sphere. Therefore, the distance between the atoms is mapped to $e_{RBF}^{(ji)} = f_c(|\vec{r}_{ji}|)\phi_{RBF}(|\vec{r}_{ji}|)$. Then a self-interaction layer $\varphi_{RBF} : R^{nb} \rightarrow R^{nh}$ is used to combine the coefficients of the radial basis functions with each other, $nh$ is the number of the hidden features of the distance, $nh = 128$ in this work. Instead of using any artificial features of the atoms, the features of the atoms are initialized by a trainable embedding layer $g : R^1 \rightarrow R^H$ based on the atomic number $Z_i$: $h_i^{(0)} = g(Z_i)$. The features of the atoms and the distances are aggregated into the edge message embeddings by the following formula:

$$m_{ji}^{(0)} = \sigma\left[W\left(g(Z_i) \oplus g(Z_j) \oplus \varphi_{RBF}\left(e_{RBF}^{(ji)}\right)\right) + b\right], \quad (13)$$

where $\sigma$ is the activation function, $\oplus$ denotes concatenation, and $W$ and $b$ are the weight matrix and bias respectively.

The triplet embedding block is shown in Fig. 2(b). The triplet message is embedded based on the distances $|\vec{r}_{kj}|$ and $|\vec{r}_{ji}|$ between node $j$ and its neighbor nodes $k$ and $i$, the angle $\alpha_{kji}$ between $\vec{r}_{kj}$ and $\vec{r}_{ji}$, and the atomic numbers $Z_k$, $Z_j$ and $Z_i$. The spherical Bessel functions are used as the two-dimensional joint bases of the distances and the angles in the triplets. The spherical Bessel function $\phi_{SBF} : (R^1, R^1) \rightarrow R^{N_{SRBF} \cdot N_{SHBF}}$ is used to expand $|\vec{r}_{kj}|$ and angle $\alpha_{kji}$ in a two-dimensional space:

$$\phi_{SBF}(|\vec{r}_{kj}|, \alpha_{kji}) = \sqrt{\frac{2}{r_c^3 j_{l+1}^2(z_{ln})}} j_l\left(\frac{z_{ln}}{r_c}|\vec{r}_{kj}|\right) Y_l^0(\alpha_{kji}), \quad (14)$$

where $j_l$ is the first kind spherical Bessel function of order $l$, and $z_{ln}$ is its $n$th root, $0 \leq l \leq N_{SHBF} - 1$, $1 \leq n \leq N_{SRBF}$. The expansion coefficients of the spherical Bessel function are multiplied by the cosine cutoff function to obtain the two-dimensional joint basis representation of $|\vec{r}_{kj}|$ and $\alpha_{kji}$: $a_{SBF}^{(kj,ji)} = f_c(|\vec{r}_{kj}|)\phi_{SBF}(|\vec{r}_{kj}|, \alpha_{kji})$. Then the coefficients of the basis functions are linearly combined through a self-interaction layer $\varphi_{SBF}: R^{N_{SRBF} \cdot N_{SHBF}} \to R^{ntri}$, $ntri$ is the number of hidden features of the triplets, $ntri = 128$ in this work. According to $a_{SBF}^{(kj,ji)}$, $e_{RBF}^{(ji)}$ and atomic numbers $Z_k$, $Z_j$ and $Z_i$, the embedding of the triplet can be constructed in the following formula:

$$m_{kji}^{(0)} = \sigma\left[W\left(g(Z_i) \oplus g(Z_j) \oplus g(Z_k) \oplus \varphi_{SBF}\left(a_{SBF}^{(kj,ji)}\right) \oplus \varphi_{RBF}\left(e_{RBF}^{(ji)}\right)\right) + b\right]. \quad (15)$$

**Update block**. The embeddings of edge messages and triplet messages are updated through multiple edge update blocks and triplet update blocks. The goal of ETGNN is to achieve the prediction of the tensor properties, so the edge update block adopts the original interaction layer in DimeNet++ to ensure that the new network has the same accuracy as the original one. The triplet update block uses the directional message passing mechanism similar to that in the edge update block. The update of the features of the triplets can be expressed by the following formula:

$$m_{kji}^{(t+1)} = f_{update}\left(m_{kji}^{(t)}, f_{int}\left(m_{kj}^{(t)}, m_{ji}^{(t)}, e_{RBF}^{(kj)}, e_{RBF}^{(ji)}, a_{SBF}^{(kj,ji)}\right)\right), \quad (16)$$

where $f_{update}$ and $f_{int}$ are update function and interaction function respectively, which are implemented by neural networks. The hidden features of the triplets $m_{kji}^{(t)}$ are updated using the directional edge messages $m_{kj}^{(t)}$ and $m_{ji}^{(t)}$ at the same stage. The angle $\alpha_{kji}$ between the directional edge messages $m_{kj}^{(t)}$ and $m_{ji}^{(t)}$ is also explicitly embedded in the spherical Bessel function $a_{SBF}^{(kj,ji)}$ to update $m_{kji}^{(t)}$.

## IV. RESULTS

### A. Prediction of force vectors

We trained ETGNN using the forces of the molecular dynamics (MD) trajectories of the extended systems provided in Ref. 26. The extended systems consist of the supercells of inorganic and organic crystals and also contain clusters and two-dimensional materials. The organic polymer pyridin and bulk $TiO_2$ contain two and three phases respectively, which can test the prediction accuracy of the model for different phases. High entropy alloy (HEA) CoCrFeMnNi contains five metal elements and each element is randomly arranged in the crystal, which is a great challenge to the model. DeepPot-SE[26] and DeePMD[14] are two representative neural network potential energy surface (PES) models, which calculate the forces by using the gradients of the potential energy: $\vec{F}_i = -\nabla_i E$. For each sub-systems, 90% randomly selected structures are used for training, and the remaining 10% are used for testing[26]. As can be seen from Table 1, the ETGNN model achieved lower RMSEs than DeepPot-SE and DeePMD on most of the sub-systems. Since training on these large extended systems is very time-consuming, the results of ETGNN listed in Table 1 were obtained within a relatively small number of training epochs but have already shown the high prediction accuracy of ETGNN for force vectors.

**Table 1.** Comparison of the root mean square errors (RMSEs) of the predicted forces for the extended systems by ETGNN, DeepPot-Se and DeePMD. The lowest RMSE among the considered models for each subsystem is displayed in bold. The unit is meV/Å.

| System | Sub-system | DeepPot-Se | DeepPMD | ETGNN |
|---|---|---|---|---|
| Bulk Cu | FCC solid | 90 | 90 | **89** |
| Bulk Ge | Diamond solid | 38 | 35 | **15** |
| Bulk Si | Diamond solid | 36 | 31 | **8** |
| Bulk $Al_2O_3$ | Trigonal solid | 49 | 55 | **36** |
| Bulk $C_5H_5N$ | Pyridine-I | **25** | **25** | 35 |
|  | Pyridine-II | **39** | **39** | 51 |
| Bulk $TiO_2$ | Rutile | 137 | 163 | **69** |
|  | Anatase | 181 | 216 | **90** |
|  | Brookite | 94 | 109 | **51** |
| $MoS_2$+Pt | $MoS_2$ slab | **23** | 34 | 30 |
|  | bulk Pt | 84 | 226 | **57** |
|  | Pt surface | 105 | 187 | **51** |
|  | Pt cluster | 201 | 255 | **115** |
|  | Pt on $MoS_2$ | **94** | 127 | 109 |
| CoCrFeMnNi HEA | rand. occ. I | 394 | 481 | **373** |
|  | rand. occ. II | 410 | 576 | **384** |

**B. Prediction of the BECs and DL tensors**

Taking the BECs of the atoms and the DL tensors $\varepsilon_\infty$ (electronic contribution) of the crystals as examples, we tested the prediction accuracy of ETGNN for the second-order tensors. DFPT calculations were performed on 3190 $Al_2O_3$, 3992 $SiO_2$ and 10000 $HfO_2$ random perturbed structures. The perturbed structures of $HfO_2$ consist of 5000 structures with 6-coordinated Hf atom and 5000 structures with 8-coordinated Hf atom. The 60%, 20% and 20% of the calculated BECs and the DL tensors were used as the training, validation and test set, respectively. Fig. 3(a−c) show the comparison of the DFT calculated BECs and the ETGNN predicted BECs on the test datasets of the perturbed structures. As can be seen from Fig. 3(a−c), each component of the BECs predicted by ETGNN is very close to the corresponding component of the BECs calculated by DFT. The mean absolute errors (MAEs) of the components of the predicted BECs on $Al_2O_3$, $SiO_2$ and $HfO_2$ test sets are only $1.18 \times 10^{-2}$ $e$, $4.28 \times 10^{-3}$ $e$ and $2.57 \times 10^{-3}$ $e$, respectively. Fig. 3(d−f) show the comparison of the DFT calculated $\varepsilon_\infty$ and the ETGNN predicted $\varepsilon_\infty$ on the test datasets of the perturbed structures. The MAEs of the components of the predicted $\varepsilon_\infty$ on $Al_2O_3$, $SiO_2$ and $HfO_2$ test sets are only $5.54 \times 10^{-3}$, $1.98 \times 10^{-3}$ and $2.86 \times 10^{-3}$, respectively. The tensor prediction test on the three perturbed structures indicates that invariant GNN can achieve high prediction accuracy in principle by using the edge-based tensor expressions proposed in this work.

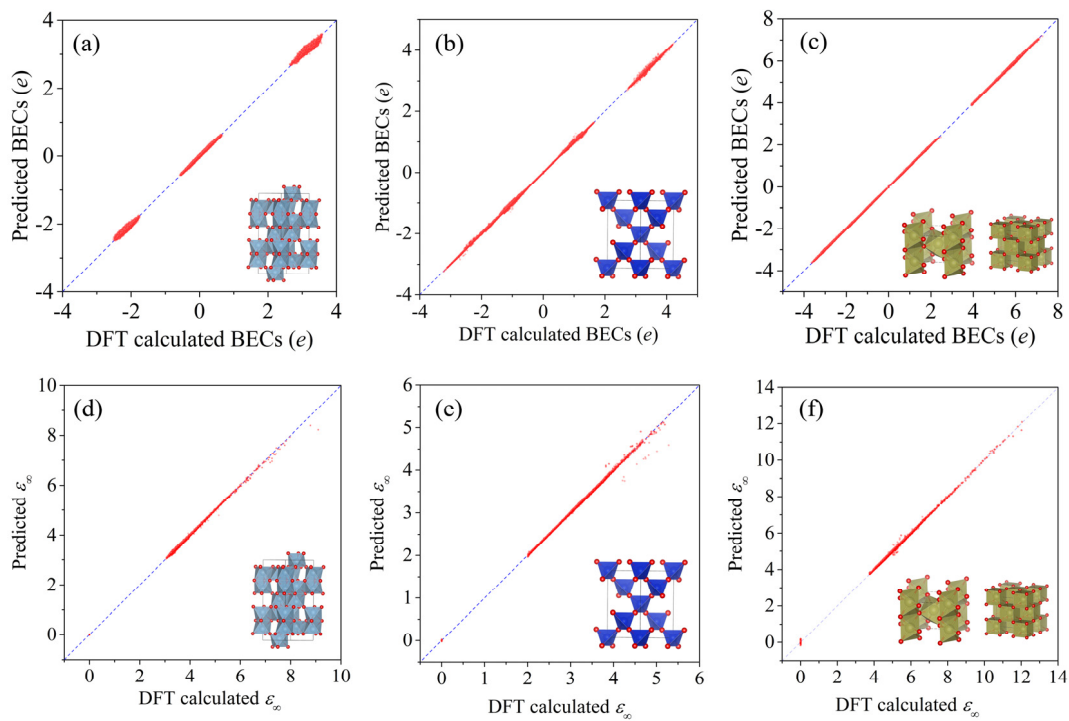

**Figure 3**: The prediction results of ETGNN on the BECs and DL tensors of the testing perturbed structure dataset and the testing JARVIS-DFT dataset. Comparison of ETGNN predicted BECs and the DFT calculated BECs on the test dataset of the random perturbed structures of (a) $Al_2O_3$, (b) $SiO_2$ and (c) $HfO_2$. Comparison of ETGNN predicted DL tensors and the DFT calculated DL tensors on the test dataset of the random perturbed structures of (d) $Al_2O_3$, (e) $SiO_2$ and (f) $HfO_2$.

ETGNN has also been tested on the BECs and DL tensors $\varepsilon_\infty$ of about 5000 non-metallic materials in the JARVIS-DFT database[27]. This data set contains more than 80 elements in the periodic table. The tensor properties in this dataset are all calculated by DFPT. Different from the dataset generated by the random perturbed structures, the structures containing some rare earth elements in this dataset may have large calculation errors, because the pseudopotentials are often hard to accurately describe the complex electronic structures of the rare earth elements. This will affect the prediction performance of the ML methods. For the JARVIS-DFT data set, the 70%, 20% and 10% of the calculated values were used as the training, validation and test set, respectively. The MAE of ETGNN on the test set of BECs still reached a low value of 0.061 *e*. This high prediction accuracy means that ETGNN can be used as a potential method to calculate BECs with much faster speed than DFT. Ref. 27 has trained a machine learning model based on the classical force-field inspired descriptors (CFID) for the maximum BECs in the JARVIS-DFT dataset. The MAE of CFID model is 0.6 *e*.[27] The performance of the traditional ML models based on manual features is generally thought to be better than deep ML models on small data sets. Instead of using any manual features, ETGNN learns the features of each element through training. For comparison with the traditional ML models, ETGNN was used to fit the maximum BECs in the JARVIS-DFT dataset. For the maximum BECs, the MAE of the ETGNN model on the test set is 0.2 *e*, which is only one-third of that of the feature-based ML model used in the ref. 27. This shows that ETGNN can also outperform the ML model based on artificial features on the small training set. It is reported that large deviations between the experimental and the calculated dielectric constants can be seen for approximately 20% of the compounds[28]. Thus predicting the dielectric tensors in various crystals is a huge challenge. The ETGNN regression model for the dielectric tensors was successfully trained and achieved an MAE of 0.218 on the test dataset.

## V. Conclusions

We have proposed a tensor prediction framework that constructs equivariant atomic tensors using the edge directions in the atomic local environment. The tensor properties of a crystal can then be regarded as the average of all the atomic tensor contributions in the crystal. Based on this idea, we have given the edge-based expansions of force vectors, BECs, DL and PZ tensors. These expansions are rotationally equivalent, while the coefficients in the expansions are invariant and can be predicted by various invariant GNNs. We designed ETGNN based on invariant MPNN to verify our tensor prediction method. For the force prediction on the extended systems, ETGNN outperforms DeePMD and DeePot-SE on most of the systems. ETGNN was used to predict the BECs and DL tensors of the perturbed structures of $Al_2O_3$, $SiO_2$ and $HfO_2$ and achieved very high accuracy. The evaluation of ETGNN on these perturbed structures indicates that the tensor expansions proposed in this work are valid. Besides, we have successfully trained the regression models of ETGNN for predicting the BECs and DL tensors of about 5000 non-metallic materials containing more than 80 elements. Taking into account the errors of the tensors calculated by DFPT, the mean absolute error of the predicted tensors in the dataset is relatively low. This shows that our model has powerful representation and generalization ability.

## Acknowledgments

This work is supported by NSFC (11825403, 11991061).

## Code availability

The code of ETGNN will be publicly available after the manuscript is accepted.

## Appendix A

**Proof of equivariance of tensor expressions.** We can prove that the edge-based tensor expansions proposed in this work are equivariant with respect to a rotation matrix $R \in \mathbb{R}^{3 \times 3}$. When the crystal structure is rotated by $R$, the basis vectors and basis tensors are rotated according to the following formulas:

$$R\vec{e}_{kj} = \vec{e}_{kj}{}' \tag{A.1}$$

$$R(\vec{e}_{kj} \otimes \vec{e}_{kj})R^{\mathrm{T}} = R\vec{e}_{kj} \otimes R\vec{e}_{kj} = \vec{e}_{kj}{}' \otimes \vec{e}_{kj}{}' \tag{A.2}$$

$$R(\vec{e}_{ji} \otimes \vec{e}_{ji})R^{\mathrm{T}} = R\vec{e}_{ji} \otimes R\vec{e}_{ji} = \vec{e}_{ji}{}' \otimes \vec{e}_{ji}{}' \tag{A.3}$$

The force $\vec{F}_j$ and tensor $T_j$ are also rotated:

$$\vec{F}_j{}' = R\vec{F}_j = \sum_{k \in N(j)} f_{kj} R\vec{e}_{kj} = \sum_{k \in N(j)} f_{kj} \vec{e}_{kj}{}' \tag{A.4}$$

$$T_j{}' = RT_j R^{\mathrm{T}} = \sum_{k \in N(j)} t_{kj} R(\vec{e}_{kj} \otimes \vec{e}_{kj})R^{\mathrm{T}} + \sum_{\substack{k,i \in N(j) \\ k \neq i}} t_{kji} R(\vec{e}_{kj} \otimes \vec{e}_{ji})R^{\mathrm{T}}$$

$$= \sum_{k \in N(j)} t_{kj} \vec{e}_{kj}{}' \otimes \vec{e}_{kj}{}' + \sum_{\substack{k,i \in N(j) \\ k \neq i}} t_{kji} \vec{e}_{kj}{}' \otimes \vec{e}_{ji}{}' \tag{A.5}$$

It can be seen from (A.4) and (A.5) that the force $\vec{F}_j$ and tensor $T_j$ are equivariant with respect to the rotation of the crystal.

**Proof of the invariance of the loss function of the tensors.** The Euclidean distance between the predicted tensor and the target tensor is used as the loss function in the present work. The loss functions of the forces and the tensors are as follows:

$$Loss_{\vec{F}} = \frac{1}{N} \sum_{i=1}^{N} \left\| \vec{F}_{i,predict} - \vec{F}_{i,target} \right\| \tag{A.6}$$

$$Loss_T = \frac{1}{N} \sum_{i=1}^{N} \left\| T_{i,predict} - T_{i,target} \right\| \tag{A.7}$$

where $N$ is the number of forces or tensors contained in each mini-batch. It is shown below that these loss functions are invariant with respect to the rotation $R$ of the crystal:

$$Loss_{R\vec{F}} = \frac{1}{N} \sum_{i=1}^{N} \left\| R\left(\vec{F}_{i,predict} - \vec{F}_{i,target}\right) \right\|$$

$$= \frac{1}{N} \sum_{i=1}^{N} \left[ R\left(\vec{F}_{i,predict} - \vec{F}_{i,target}\right) \right]^{\mathrm{T}} \left[ R\left(\vec{F}_{i,predict} - \vec{F}_{i,target}\right) \right]$$

$$= \frac{1}{N}\sum_{i=1}^{N}\left(\vec{F}_{i,predict}-\vec{F}_{i,target}\right)^{\mathrm{T}}R^{\mathrm{T}}R\left(\vec{F}_{i,predict}-\vec{F}_{i,target}\right)$$

$$= \frac{1}{N}\sum_{i=1}^{N}\left(\vec{F}_{i,predict}-\vec{F}_{i,target}\right)^{\mathrm{T}}\left(\vec{F}_{i,predict}-\vec{F}_{i,target}\right)$$

$$= \frac{1}{N}\sum_{i=1}^{N}\left\|\vec{F}_{i,predict}-\vec{F}_{i,target}\right\|$$

$$= Loss_{\vec{F}} \tag{A.8}$$

First, we can write formula (A.7) as:

$$Loss_T = \frac{1}{N}\sum_{i=1}^{N}\left\|T_{i,predict}-T_{i,target}\right\|$$

$$= \frac{1}{N}\sum_{i=1}^{N}tr\left[\left(T_{i,predict}-T_{i,target}\right)\left(T_{i,predict}-T_{i,target}\right)^{\mathrm{T}}\right] \tag{A.9}$$

Under the rotation matrix $R$, the loss function of the tensor $T$ can be written as:

$$Loss_{RTR^{\mathrm{T}}} = \frac{1}{N}\sum_{i=1}^{N}tr\left[\left[R\left(T_{i,predict}-T_{i,target}\right)R^{T}\right]\left[R\left(T_{i,predict}-T_{i,target}\right)R^{T}\right]^{\mathrm{T}}\right]$$

$$= \frac{1}{N}\sum_{i=1}^{N}tr\left[R\left(T_{i,predict}-T_{i,target}\right)R^{T}R\left(T_{i,predict}-T_{i,target}\right)^{T}R^{T}\right]$$

$$= \frac{1}{N}\sum_{i=1}^{N}tr\left[R\left(T_{i,predict}-T_{i,target}\right)\left(T_{i,predict}-T_{i,target}\right)^{T}R^{T}\right]$$

$$= \frac{1}{N}\sum_{i=1}^{N}tr\left[R^{T}R\left(T_{i,predict}-T_{i,target}\right)\left(T_{i,predict}-T_{i,target}\right)^{T}\right]$$

$$= \frac{1}{N}\sum_{i=1}^{N}tr\left[\left(T_{i,predict}-T_{i,target}\right)\left(T_{i,predict}-T_{i,target}\right)^{T}\right]$$

$$= Loss_T \tag{A.10}$$

Eq. (A.8) and (A.10) show that the Euclidean distance loss functions are rotationally equivariant. The following linear algebra formulas are used in the above proof process:

$$\left(AB\right)^{\mathrm{T}} = B^{\mathrm{T}}A^{\mathrm{T}} \tag{A.11}$$

$$\|A\| = tr\left(AA^{\mathrm{T}}\right) \tag{A.12}$$

$$tr(ABC) = tr(CAB) = tr(BCA) \qquad (A.13)$$